\documentstyle[12pt,aps,epsf]{revtex}

\input{epsf}
\begin{document}

\title{ The resonant behaviour of the Faraday
 rotation in a  medium with linear birefringence}
\author{Kozlov G.G.}

\maketitle
 \hskip20pt
All-Russia Research Center "Vavilov State Optical Institute"
\vskip20pt
\hskip100pt {\it e}-mail:  gkozlov@photonics.phys.spbu.ru
\vskip20pt
\begin{abstract}
It is shown that  the monochromatic optical wave propagating
through the medium with linear birefringence in presence of a
signal electromagnetic wave (whose wavelength is equal to the polarization
beats length), displays Faraday rotation having the
frequency of the signal wave and unsuppressed by linear birefringence.
The effect is resonant with respect to the frequency of a signal wave.
The "sharpness" of the resonance is defined by
 length of the birefringent medium.
\end{abstract}

\section{Introduction}

In the present paper we suggest the effect of resonant behaviour
of the Faraday rotation of light propagating through a medium
with linear birefringence. Due to the universality of the effect suggested
the resonant frequency of the magnetic field has nothing to do
with resonances of the susceptibility of the medium and completely
defined by the value of linear birefringence of the medium in the
transparency spectral region.
 In this section we describe the essence of the effect. For this reason we
 briefly  remind some statements of the polarization optics.
The optical wave electrical displacement vector
in the medium runs over the ellipse whose plane is
perpendicular to the direction of wave propagation.
The parameters of this ellipse  define the wave polarization.
The case of ellipse  squeezed to a line corresponds to linear
polarization, the case of circle corresponds to circular polarization.
The relative amplitudes and phases of the projections of this motion
with respect to some fixed coordinate system  completely define
the polarization state. These projections can be described by Jones vector:
\begin{equation}
|\bf{ \xi}>=\left(\matrix{\cos\phi\exp i\delta \cr\sin\phi}\right)
\end{equation}
where $\cos\phi$ and $\sin\phi$ -- relative amplitudes,
 and $\delta$ -- relative phase of the projections.
For an arbitrary  direction in anisotropic medium one can specify
two linearly independent monochromatic waves -- normal modes --
propagating with no change of the polarization state. The  normal
modes Jones vectors $|{\bf\xi}_\pm>$ are the eigen vectors of the inverse permittivity
tensor projected on the plane perpendicular to the wave propagation direction.
The corresponding eigen values define the normal
 modes refractive indexes $n_\pm$\cite{Landau}.
Below we consider the transparent anisotropic medium for which
the matrix of this tensor is hermitian with its real symmetric
part describing linear birefringence and imaginary antisymmetric
part describing circular birefringence.
For the problems of polarization optics not the refractive indexes
of the normal modes themselves are of prime importance but their
difference. Omitting some detailes which are not important for us
we can introduce tensor $\hat{\eta}$ whose eigen values reproduce
the deviations of refractive indexes of normal modes from their
main (or average) value (which we denote by $n$) and whose eigen vectors
reproduce the Jones vectors of the normal modes.
Real symmetric part of this tensor describe the linear
birefringence and imaginary antisymmetric part describe the
circular birefringence. In the coordinate system where
the real part of this tensor is diagonal it can be
always written in the following convenient form:
\begin{equation}
2\hat{\eta} \equiv h_0\hat{\sigma}_z+h_1\hat{\sigma}_y
\end{equation}

 where $\hat{\sigma}_z$ and $\hat{\sigma}_y$ -- Pauli spin-matrices:

\begin{equation}
\hat{\sigma}_z\equiv{1\over 2}\left(\matrix{1&0\cr
0&-1}\right),\hskip10mm
\hat{\sigma}_y\equiv{1\over 2}\left(\matrix{0&-i\cr
i&0}\right)
\end{equation}
In accordance with the above we call coefficient $h_0$  {\it linear birefringence}
and  coefficient $h_1$ we call {\it gyration}.
The arbitrary polarized incident light can be always reproduced as a
linear combination of the normal modes.
Due to the fact that the propagation velocities of the normal modes  differ
from each other the polarization state will display the periodical changes
along the propagation direction. The corresponding spatial period $\Lambda_{beats}$
 called the polarization beats length and can be calculated as:
   \begin{equation}
   \Lambda_{beats}={\Lambda\over |n_--n_+|},
   \end{equation}
   Here $\Lambda$ -- is the vacuum wavelength of the incident light.
For the case of the media with linear birefringence the
normal modes are linearly polarized in two orthogonal
directions and the polarization beats represent  the
periodical changes of ellipticity. For the case of
circular birefringence (which appear for example in the isotropic
medium affected by the  magnetic field) the normal modes are the circularly
polarized waves with different  direction of rotation.
In this case (if the incident wave is linearly polarized)
the polarization beats represent the rotation of the
polarization plane. Below we consider $h_0, h_1<< n\equiv(n_++n_-)/2$,
  which is almost always the case for the actual media.
 Now we are ready to pass to our effect. Suppose that  the linearly polarized
 wave falls on the medium having length $L$ with linear
 birefringence $h_0$. Suppose that  this wave  has
   one of  eigen polarizations of the medium. This wave
 will pass through the medium with no change of  polarization state.
 Now let us switch on  the weak longitudinal magnetic field
 which gives rise to gyration $h_1<<h_0$.
If the linear birefringence is zero, the polarization
plane would rotate by an angle $\pi L h_1/\Lambda$. Nevertheless it is known that
if $h_1<<h_0$ (weak gyration) and $L>>\Lambda_{beats}=2\Lambda/h_0$
(the medium length is greater than polarization beats spatial period)
the above rotation is considerably suppressed \cite{Chetkin}.
 This suppression can be removed as follows \cite{Chetkin,Zap}.
Suppose that the magnetic field is periodically
 varying in space and has the spatial
period $2\Lambda/h_0$  i.e. equal to that of polarization beats.
For example let gyration be the following function of
coordinate: $\sim h_1\cos (\pi h_0 x/\Lambda)$.
 Then (under some additional conditions \cite{Zap}) the
 polarization plane of the beam on exit of the medium rotate
by an angle $\pi L h_1/\Lambda$, equal to that for the case of
zero linear birefringence. It is wellknown effect of compensation
of  linear birefringence \cite{Chetkin,Zap}.
       {\it In this paper we suggest the effect of  similar
       compensation by means of magnetic field which is constant
       in space (along the beam propagation direction) but varying in time.}
Let us introduce the incident plane wave as a sequence
of pulses, each of which contain several spatial periods of the
incident wave $\Lambda$. One of these pulses is depicted on
Fig.1 by a bold line. The temporal dependence of  coordinate $x(t)$
of an arbitrary pulse has the form:
       \begin{equation}
       x(t)=x_0+{c\over n} t
       \end{equation}

Here $x_0$ -- is the pulse coordinate at $t=0$, $n$ -- the
 main value of  refractive index
of the medium, $c$ -- the speed of light.
 Pulse can be identified by the value of $x_0$ and for
this reason we denote  the pulse
whose coordinate is $x_0$ at $t=0$ by a term "pulse $x_0$".
 Now let us affect our medium by a time-dependent magnetic
field directed along the $x$- axis.
If we denote by $\omega$ the frequency of this magnetic field,
then this field gives rise to a gyration which can be written in the form:
 \begin{equation}
 h=h_{1}\cos(\omega t)
 \end{equation}

It is easy to see that from the point of view of the pulse $x_0$
the gyration depends on $x$. For this to see one should
express $t$ from (5) and substitute it in (6):

\begin{equation}
h=h_{1}\cos\bigg[\omega n {x-x_0\over c}\bigg]
\end{equation}

By changing the frequency of magnetic field $\omega$ one can
adjust the spatial period of gyration (7) to be equal to
that of polarization beats related to the linear
birefringence of the medium. This leads to the
following relationship  for this frequency:

 \begin{equation}
 \Lambda_{beats}={2\Lambda\over h_0}={2\pi c\over n\omega}\equiv{\lambda\over n}
 \end{equation}
 $$
 \omega={\pi c h_0\over n \Lambda}
 $$
Here $\lambda$ is the vacuum wavelength of the
electromagnetic wave having the frequency $\omega$.
 So we see that the linear birefringence is compensated for
the pulse $x_0$ and this pulse
(under some additional conditions which we consider below)
will display unsuppressed Faraday rotation $\pi L h_1/\Lambda$.
This effect is resonant with respect to the deviation of  frequency
of the magnetic field from the value defined by (8). In the next
section we present the quantitative calculation and show that
the above  resonance is accompanied by oscillations of the
polarization plane with frequency $\omega$.

\section{Calculation}

In this section we present a calculation of the above effect.
It is convenient to perform calculations by means of the {\it quasispin}
method - the method of polarization optics based on the
formal similarity of the dynamics of  quantum two-level system
and the dynamics of  polarization of light propagating in the anisotropic
medium. Let us remind briefly the essence of this method.
The polarization state  is described by the vector of quasispin ${\bf S}$
whose components $S_i$ expressed via the Pauli spin matrices $\hat{\sigma}_i$
and Jones vector (1) as follows:

  \begin{equation}
  S_i\equiv<\xi|\hat{\sigma_i}|\xi>\hskip10mm i=1,2,3
  \end{equation}

While propagating through the medium the polarization of the
beam can change and consequently the direction of quasispin ${\bf S}$
depends on $x$-coordinate along the propagation direction. This
dependence can be described by the  Bloch equation:

 \begin{equation}
 d{\bf S}/dD=[{\bf H, S}]
 \end{equation}
 The dimensionless coordinate $D$  plays the role of "time" and
 can be expressed in terms of conventional coordinate $x$ as:
$D\equiv 2\pi x/\Lambda$.
 The effective "magnetic field" vector can be expressed via the
   linear birefringence and gyration as follows:
 ${\bf H}\equiv(0,h_1,h_0)$.
The quantities $h_1$ and $h_0$ can depend on $D$. Below we
deal with the gyration having the sinusoidal dependence on $D$.
This dependence can be removed  by passing to the
rotating frame in the Bloch equation (10)\cite{Zap}.
 Let us now turn to the problem we are interested in.

Consider the monochromatic beam having the vacuum wavelength
$\Lambda$ propagating in a medium with linear birefringence $h_0$
and with average refractive index $n$ (Fig.1) We are interested
in the dynamics of the polarization of this wave and we call this
wave the {\it  optical} wave.
 Let the other monochromatic wave (which will be
  hereinafter referred to as {\it signal wave})
be falling at the medium at an angle $\beta$ so the vector
of the magnetic field in this wave has the $x$- component
 $\sim\cos\beta$.
The magnetic field of the signal wave give rise to the gyration in
the medium and the polarization of optical wave acquire the increment
which we are interested in. Let $\lambda$
 be  the vacuum wavelength of the signal wave, $n_1$
be the refractive index of the enviroment, $L$  be the medium
length. Let the thickness of the medium be much smaller
than $\lambda$ so the magnetic field of the signal wave
is homogeneous within our medium. The setup described
corresponds to the case of the planar waveguide with the
optical wave propagating inside it  and irradiated by the signal wave
falling at an angle $\beta$ to the plane of the waveguide.
It is easy to see that the distribution of gyration in the
medium, produced by a signal wave, can be written as:
 \begin{equation}
 h=h_1\cos\bigg[\omega t -x {2\pi\over\lambda} n_1\sin\beta\bigg]
 \end{equation}
 $$
 h_1\equiv 2 H_0 V\cos\beta
 $$
 where $V$ -- the Verdet constant of the medium, $H_0$ --
 the amplitude of the magnetic field in the signal wave, $\omega$ --
 the frequency of field oscillations in it.
Now let us consider this distribution from the point of view
of the pulse $x_0$ of the optical wave -- in the same way it was
done in the Introduction. For this reason express $t$ from
(5) and substitute it in (11). Passing to the dimensionless
coordinate one can obtain that from the pulse $x_0$ point of view
 the following spatial  distribution of the gyration take place:

   \begin{equation}
   h(D)=h_1\cos\bigg[\Omega D + \phi_0\bigg]
   \end{equation}
$$
\Omega\equiv{\Lambda\over\lambda}(n-n_1\sin\beta),\hskip10mm
\phi_0\equiv-{2\pi n\over\lambda} x_0
$$

If we denote the optical wave quasispin vector on the
entrance of the medium by ${\bf S}_0$, then the temporal behaviour
of the quasispin of the pulse $x_0$ can be calculated from
the Bloch equation (10) with the effective
"magnetic field" ${\bf H}=(0,h,h_0)$ and under the initial condition ${\bf S}(0)={\bf
S}_0$. The equation (10) should be solved
 within the interval $D\in[0,2\pi L/\Lambda]$.
The solution of this problem can be carried out
in a conventional way and consists of the following steps:

1. Replace the effective "magnetic field"  oscillating in $y$-direction
by  the field  rotating in $xy$ plane and having the same $y$-component.

2. Pass to the frame rotating with "frequency" $\Omega$.
 We denote the matrix of corresponding coordinate
 transformation by $\hat{R}$. The effective
 "magnetic field" in the rotating $R$ -frame has the form:
$$
\tilde{\bf{H}}=(0,h_0-\Omega,h_1).
$$

3. Perform the rotation around the $x$-axis of the $R$ -frame by an angle:
$$
 \theta=\arctan{h_0-\Omega\over h_1}.
$$
We denote the matrix of this transformation by
$\hat{T}$. In this $TR$ -frame the effective "magnetic field" has the components:
$$
  \tilde{\tilde{\bf H}}=(0,0,\tilde{\Omega}),
$$
where
$$
\tilde{\Omega}\equiv\sqrt{h_1^2+(h_0-\Omega)^2}
$$
The solution of Bloch equation in $TR$ -frame represent the
rotation around $z$-axis with "frequency" $\tilde{\Omega}$.
The corresponding matrix has the form:
$$
\hat{E}=\left(\matrix{ \cos\tilde{\Omega} D& \sin\tilde{\Omega} D& 0\cr
-\sin\tilde{\Omega} D & \cos\tilde{\Omega} D & 0\cr
0&0&1}\right),
$$
i.e., if $\tilde{\tilde{{\bf S}}}_0=\hat{T}\hat{R}\hskip1mm{\bf S}_0$
 is the quasispin of light at $D=0$ in $TR$ -frame,
 then for the arbitrary
 $D$, quasispin in $TR$ -frame $\tilde{\tilde{{\bf S}}}(D)$
 can be written as: $\tilde{\tilde{{\bf S}}}(D)=\hat{E}\tilde{\tilde{{\bf
 S}}}_0$.
 To obtain the final result one should pass back to the initial frame.
  Consequently,  the quasispin dynamics  can be calculated as:
 \begin{equation}
 {\bf S}(D)=\hat{R}^{-1}\hat{T}^{-1}\hat{E}\hat{T}\hat{R}{\bf S}_0
 \end{equation}
 The explicit expressions for the above matrices are:
 \begin{equation}
 \hat{R}=\left(\matrix{\cos(\phi_0-\Omega D)& -\sin(\phi_0-\Omega D)&0\cr
 \sin(\phi_0-\Omega D) & \cos(\phi_0-\Omega D)& 0\cr
 0&0&1}\right)
 \end{equation}

\begin{equation}
\hat{T}=\left(\matrix{1&0&0\cr
0&\sin\theta&-\cos\theta\cr
0&\cos\theta&\sin\theta}\right)
\end{equation}

To calculate the quasispin on the exit of the medium one should set:
$D=2\pi
L/\Lambda$.
Remind now that we make our calculations for the polarization of the pulse
$x_0$. This pulse goes out of the medium at $t=n(L-x_0)/c$.
 Taking into account the expression (12) we obtain that
 for calculation of the {\it temporal} dynamics of the quasispin
 on the exit of the medium one should set
\begin{equation}
 \phi_0=\omega t -{2\pi n\over\lambda}L.
\end{equation}
in formula (13).
Let us consider the simple example with the incident optical
wave having the eigen polarization of linear birefringent medium.
In this case the initial quasispin on the entrance of the medium
is directed along $z$-axis for an arbitrary pulse of the optical wave:

\begin{equation}
{\bf S}_0=(0,0,1)
\end{equation}

For this initial condition let us perform all the matrix
multiplications in (13) except the last one ($\hat{R}^{-1}$). For the
quasispin on the exit of the medium we obtain:

%\newpage
\begin{equation}
{\bf S}(D)=
\left(\matrix{\cos(\phi_0-\Omega D)& \sin(\phi_0-\Omega D)&0\cr
 -\sin(\phi_0-\Omega D) & \cos(\phi_0-\Omega D)& 0\cr
 0&0&1}\right) {\bf S}_1
 \end{equation}
$$
 {\bf S}_1\equiv
 \left(\matrix{-\cos\theta\sin(\tilde{\Omega} D)\cr
 \sin\theta\cos\theta[1-\cos(\tilde{\Omega}D)]\cr
 \cos^2\theta\cos(\tilde{\Omega}D)+\sin^2\theta}\right)
$$
Note that ${\bf S}_1$ has no temporal dependence.
It is seen that the temporal behaviour of  ${\bf S}(D)$ represent the rotation of
vector ${\bf S}_1$ around $z$-axis with  frequency $\omega$ of the signal wave.

When  $h_0=\Omega$, $\theta=0$ the polarization
 state of optical wave displays maximum change:
the quasispin vector runs over the cone with an angle $h_1 D=2\pi h_1 L/\Lambda$
what  corresponds to the twice smaller oscillations of the
polarization plane $\pi h_1 L/\Lambda$, i.e. to the Faraday rotation of the optical wave
 unsuppressed by the linear birefringence.
The direct calculation shows that the amplitude $\phi$ of the Faraday rotation
of the optical wave is:

 \begin{equation}
 \phi={1\over 2}\arcsin\sqrt{1-S_{1z}^2}
 \end{equation}
 where
\begin{equation}
S_{1z}={h_1^2\over h_1^2+(h_0-\Omega)^2}
\bigg[\cos \sqrt{h_1^2+(h_0-\Omega)^2}D +\bigg({h_0-\Omega\over
h_1}\bigg)^2\bigg]
\end{equation}
$$
D=2\pi {L\over\Lambda}
$$
One can pass the above resonance not by changing frequency $\omega$
but by changing $\beta$ (12).
The dependence of $\phi$ on $\beta$ (Fig.2) exhibit the resonant behaviour.
The "sharpness" of this resonance is $\sim L/\Lambda_{beats}$.

\newpage
\begin{figure}
\epsfxsize=400pt
\epsffile{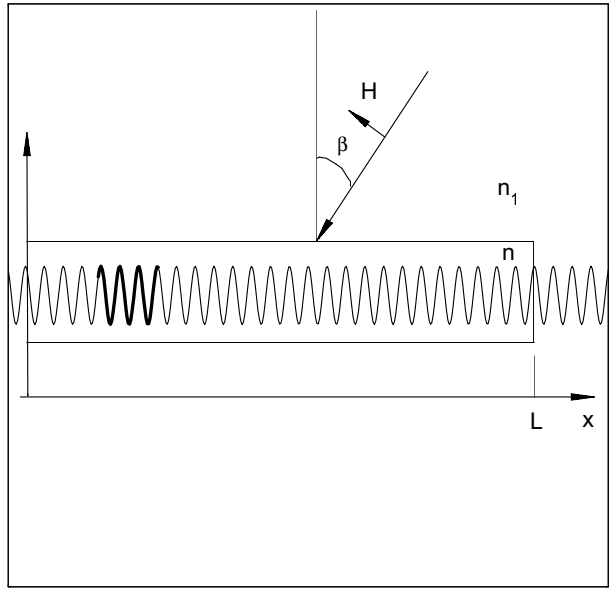}
\caption{The resonant behaviour of the Faraday
 rotation in a  medium with linear birefringence.
Horizontal rectangular -- birefringent medium,  sine
function inside -- optical wave. Bold fragment of the sine function --
one of the pulses the optical wave consist of.  Tilted arrow --
the signal wave.}
\end{figure}
\begin{figure}
\epsfxsize=400pt
\epsffile{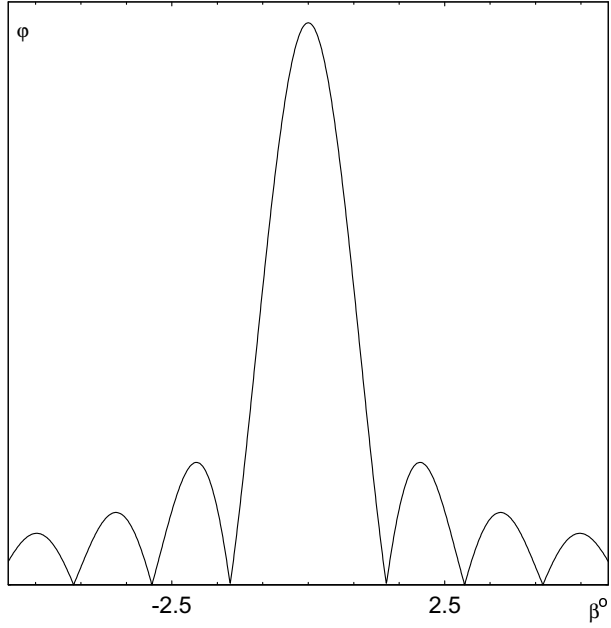}
\caption{ The dependence  of the Faraday rotation on
an incident angle $\beta$ of the signal wave for the following values of the
 parameters:
 $n=n_1=2$, $h_0=0.02$, $\Lambda=1\mu m$, $\lambda=100\mu m$, $L=3$mm.}
\end{figure}

\end{document}